\documentclass[11pt]{article}
\usepackage{moriond,epsfig}

\def\be{\begin{equation}}
\def\ee{\end{equation}}
\def\bea{\begin{eqnarray}}
\def\eea{\end{eqnarray}}

\begin{document}
\vspace*{4cm}
\title{THE ANOMALOUS $\Upsilon(1S)\pi^+ \pi^-$ AND $\Upsilon(2S)\pi^+ \pi^-$
PRODUCTION NEAR THE $\Upsilon(5S)$ RESONANCE}

\author{ K.-F.~Chen }

\address{Department of Physics, National Taiwan University, 
No. 1, Sec. 4, Roosevelt Road, Taipei, 10617 Taiwan(R.O.C)}

\maketitle\abstracts{
We report the study of $e^+e^- \to
\Upsilon(1S)\pi^+\pi^-$, $\Upsilon(2S)\pi^+\pi^-$, $\Upsilon(3S)\pi^+\pi^-$, and
$\Upsilon(1S)K^+K^-$, near the peak of the $\Upsilon(5S)$
resonance. The results are based on a
data sample of 21.7~fb$^{-1}$ collected with the Belle detector at
the KEKB $e^+e^-$ collider. 
Attributing the signals to the $\Upsilon(5S)$ resonance, 
the partial widths
$\Gamma(\Upsilon(5S)\to\Upsilon(1S)\pi^+\pi^-) = 0.59\pm0.04{\rm (stat)}\pm0.09{\rm (syst)}$ MeV and
$\Gamma(\Upsilon(5S)\to\Upsilon(2S)\pi^+\pi^-) = 0.85\pm0.07{\rm (stat)}\pm0.16{\rm (syst)}$ MeV 
are obtained from the observed cross sections. 
These values exceed by more than two orders of magnitude 
the previously measured partial widths for dipion transitions between lower $\Upsilon$ resonances.}

\section{Introduction}

The dipion transitions between $\psi$ or $\Upsilon$ levels below the open 
flavor thresholds have been successfully
described in terms of QCD multipole moments~\cite{Brown:1975dz}. 
The measurements of $\Upsilon(4S) \to \Upsilon(1S)\pi^+\pi^-$~\cite{CLEO:2007sja,Aubert:2006bm,Sokolov:2006sd}
are consistent with this picture~\cite{Simonov:2007bm}. 
However, The spectroscopy above open flavor threshold is complex, 
The recent discovery of a broad $1^{--}$ state, the $Y(4260)$,
decaying with an unexpectedly large partial width to
$J/\psi \pi^+\pi^-$~\cite{Y4260}, has brought new challenges to the interpretation of its composition, 
with ``hybrid" $c\overline{c}g$ (where $g$
is a gluon) and $c\overline{c}q\overline{q}$ (where $q\overline{q}$ is a color-octet light quark
pair) four quark state as possibilities. The observation of
a bottomonium counterpart to $Y(4260)$, which we shall
refer to as $Y_b$~\cite{Hou:2006it}, could shed further light on the structure
of such particles. The expected mass is above the $\Upsilon(4S)$.
It has been suggested that a $Y_b$ with lower mass can be
searched for by radiative return from the $\Upsilon(5S)$, and one
with higher mass through an anomalous rate of $\Upsilon(nS)\pi\pi$
events~\cite{Hou:2006it}; scaling from $\Upsilon(4S)\to\Upsilon(1S)\pi\pi$, one expects
$\Upsilon(5S)\to\Upsilon(1S)\pi\pi$ to have branching fraction $\sim 10^{-5}$.

In our studies, the rates for $\Upsilon(1S) \pi^+ \pi^-$ and $\Upsilon(2S) \pi^+ \pi^-$
are found to be much larger than the expectations from scaling the
comparable $\Upsilon(4S)$ decays to the $\Upsilon(5S)$.
Since only one center-of-mass (CM) energy is
used, one does not know whether 
these enhancements are an effect of
the $\Upsilon(5S)$ itself, or due to a nearby or overlapping $``Y_b"$ state. 
Throughout this proceeding, we use the notation $\Upsilon(10860)$ instead of
$\Upsilon(5S)$.

\section{The Analysis}

This study is described in details in the reference~\cite{Abe:2007tk}.
The $\Upsilon(10860)\to\Upsilon(nS)\pi^+\pi^-$ and
$\Upsilon(1S)K^+K^-$ final states are reconstructed using
$\Upsilon(nS) \to \mu^+\mu^-$ decays. 
Events with exactly four well-constrained charged tracks
and zero net charge are selected.
Two muons with opposite charge are selected to
form a $\Upsilon(nS)$ candidate.
The two remaining tracks are treated as pion or kaon candidates. To
suppress the background from $\mu^+\mu^-\gamma \to
\mu^+\mu^-e^+e^-$ with photon conversion, pion candidates with
positive electron identification are rejected. 
The cosine of the
opening angle between the $\pi^+$ and $\pi^-$ ($K^+$ and $K^-$)
momenta in the laboratory frame is required to be less than 0.95.
The trigger efficiency is found to be very
close to 100\% for these final states.
To reject (radiative) Bhabha and $\mu$-pair backgrounds, the data are 
required to satisfy either $\theta_{\rm max}<175^\circ$, or 2 GeV $<\sum
E_{\rm ECL} < 10$ GeV, where $\theta_{\rm max}$ is the 
maximum opening angle between any charged tracks in the CM frame,
and $\sum E_{\rm ECL}$ is the sum of the calorimeter energy.

The signal candidates are identified using the kinematic variable
$\Delta M$, defined as the difference between
$M(\mu^+\mu^-\pi^+\pi^-)$ or $M(\mu^+\mu^-K^+K^-)$ and
$M(\mu^+\mu^-)$ for pion or kaon modes. Sharp peaks are expected at
$\Delta M = M_{\Upsilon(mS)} - M_{\Upsilon(nS)}$ for $m>n$. For
$\Upsilon(10860)\to\Upsilon(nS)\pi^+\pi^-$ and
$\Upsilon(1S)K^+K^-$, signal events should be concentrated at $\Delta
M = \sqrt{s} - M_{\Upsilon(nS)}$, since a single CM energy is
used. 

\begin{figure}[b!]
\begin{center}
\includegraphics[width=9.5cm]{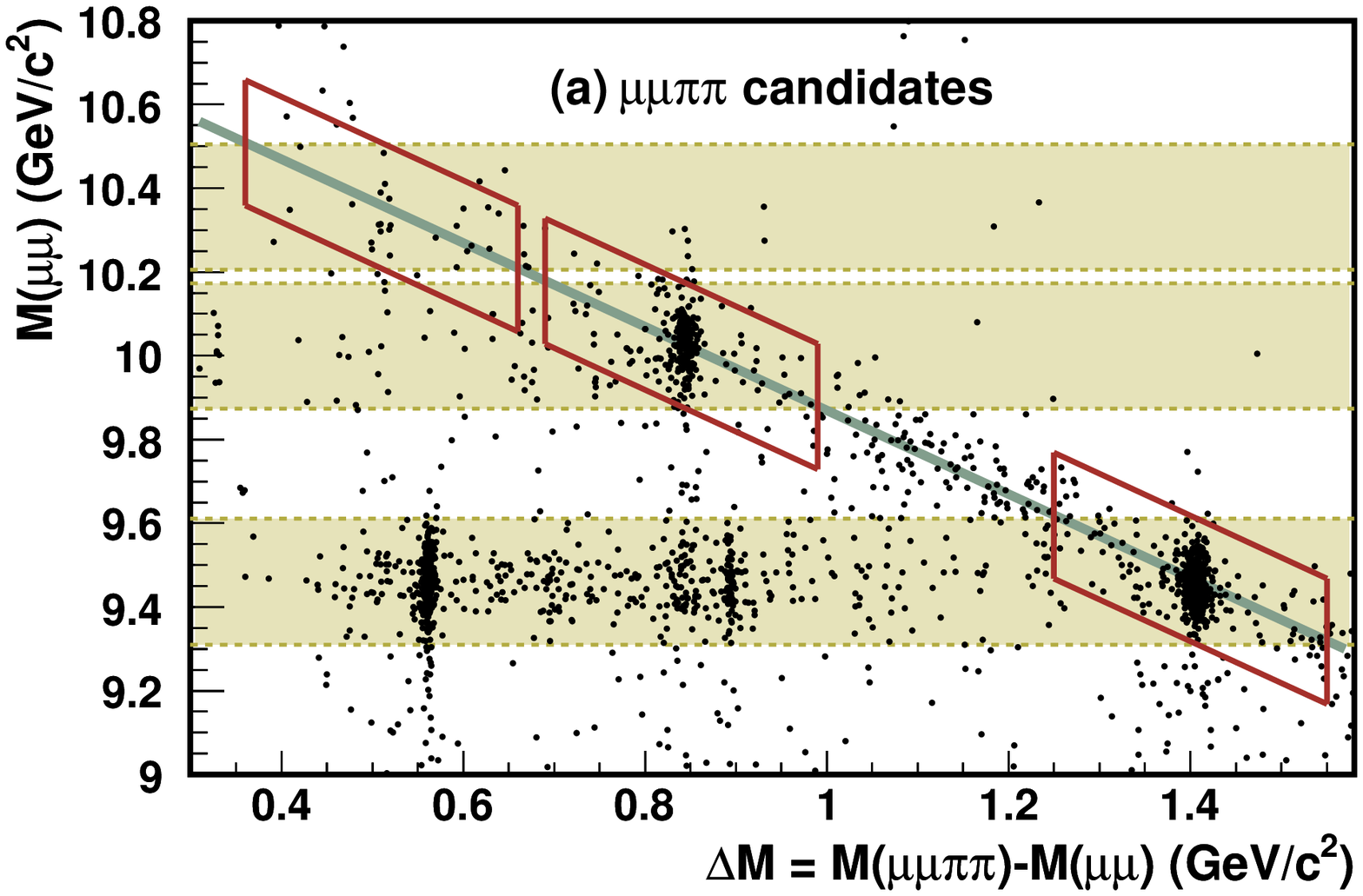}
\includegraphics[width=9.5cm]{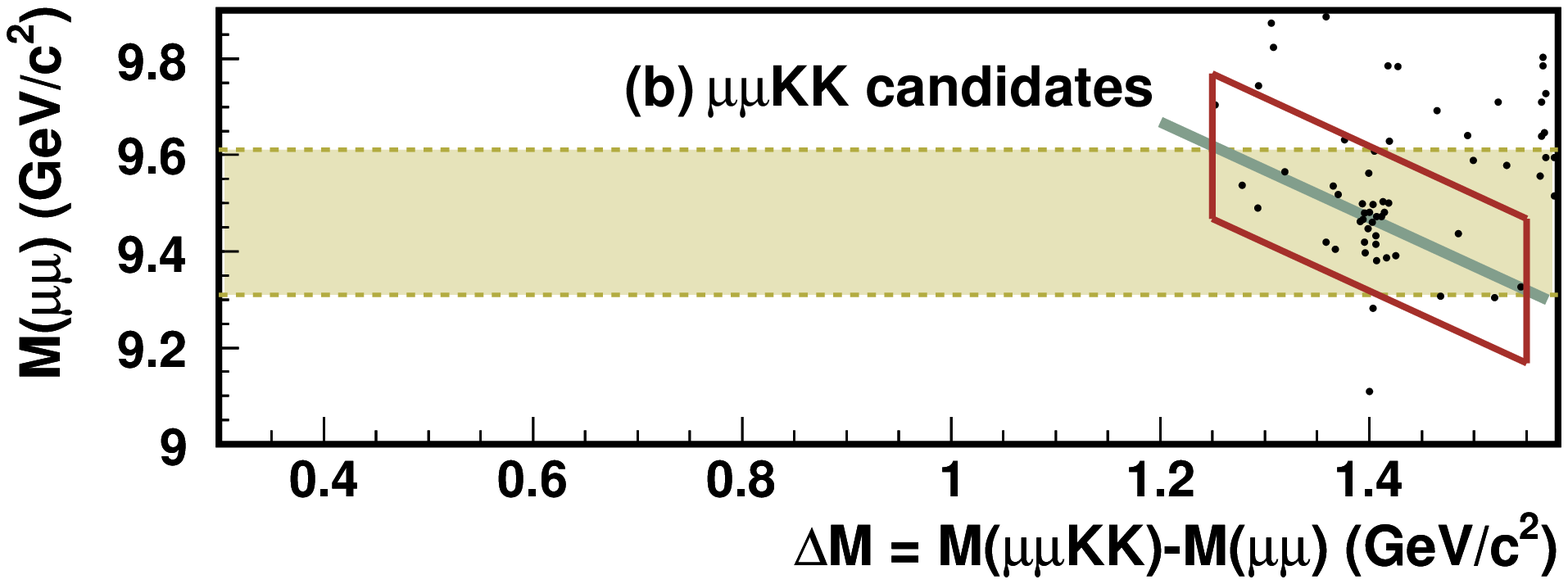}
\end{center}
\caption{Scatter plot of $M(\mu^+\mu^-)$ vs.
$\Delta M$ for the data collected at $\sqrt{s}\sim10.87$ GeV, for
(a) $\mu^+\mu^-\pi^+\pi^-$ and (b) $\mu^+\mu^-K^+K^-$ candidates.
Horizontal shaded bands correspond to $\Upsilon(1S)$,
$\Upsilon(2S)$ and $\Upsilon(3S)$ (only $\Upsilon(1S)$ for (b)),
and open boxes are the fitting regions for
$\Upsilon(10860)\to\Upsilon(nS)\pi^+\pi^-$ and
$\Upsilon(1S)K^+K^-$. The lines indicate the kinematic boundaries, 
$M(\mu^+\mu^-\pi^+\pi^-, \mu^+\mu^-K^+K^-) = \sqrt{s}$.}
 \label{fig:ym_vs_dm}
\end{figure}

Figure~\ref{fig:ym_vs_dm} shows the
two-dimensional scatter plot of $M(\mu^+\mu^-)$ vs. $\Delta M$ for
the data. Clear enhancements are observed, especially for
$\Upsilon(10860)\to\Upsilon(1S)\pi^+\pi^-$ and
$\Upsilon(2S)\pi^+\pi^-$ decays. 
The dominant background processes, $e^+e^- \to \mu^+\mu^-\gamma(\to e^+e^-)$ and $e^+e^- \to \mu^+\mu^-\pi^+\pi^-$ 
accumulate at the kinematic boundary, $M(\mu^+\mu^-\pi^+\pi^-) = \sqrt{s}$.
The events with $|M(\mu^+\mu^-\pi^+\pi^-)-\sqrt{s}|<150$ MeV
or $|M(\mu^+\mu^-K^+K^-)-\sqrt{s}|<150$ MeV are selected. The
fitting regions are defined by 1.25 GeV/$c^2$ $<\Delta M<$ 1.55
GeV/$c^2$, 0.69 GeV/$c^2$ $<\Delta M<$ 0.99 GeV/$c^2$, and 0.36
GeV/$c^2$ $<\Delta M<$ 0.66 GeV/$c^2$ for
$\Upsilon(10860)\to\Upsilon(1S)\pi^+\pi^-$,
$\Upsilon(2S)\pi^+\pi^-$, and $\Upsilon(3S)\pi^+\pi^-$,
respectively. The fitting region in $\Delta M$ for
$\Upsilon(10860)\to\Upsilon(1S)K^+K^-$ is the same as for the
$\Upsilon(1S)\pi^+\pi^-$ mode. 
The oblique fitting regions 
are selected so that the background shape is monotonic along each band.
The background distributions are verified 
using the off-resonance sample (recorded at $\sqrt{s} \sim 10.52$ GeV)~\cite{Sokolov:2006sd}.

\begin{figure}[b!]
\begin{center}
\includegraphics[width=9.5cm]{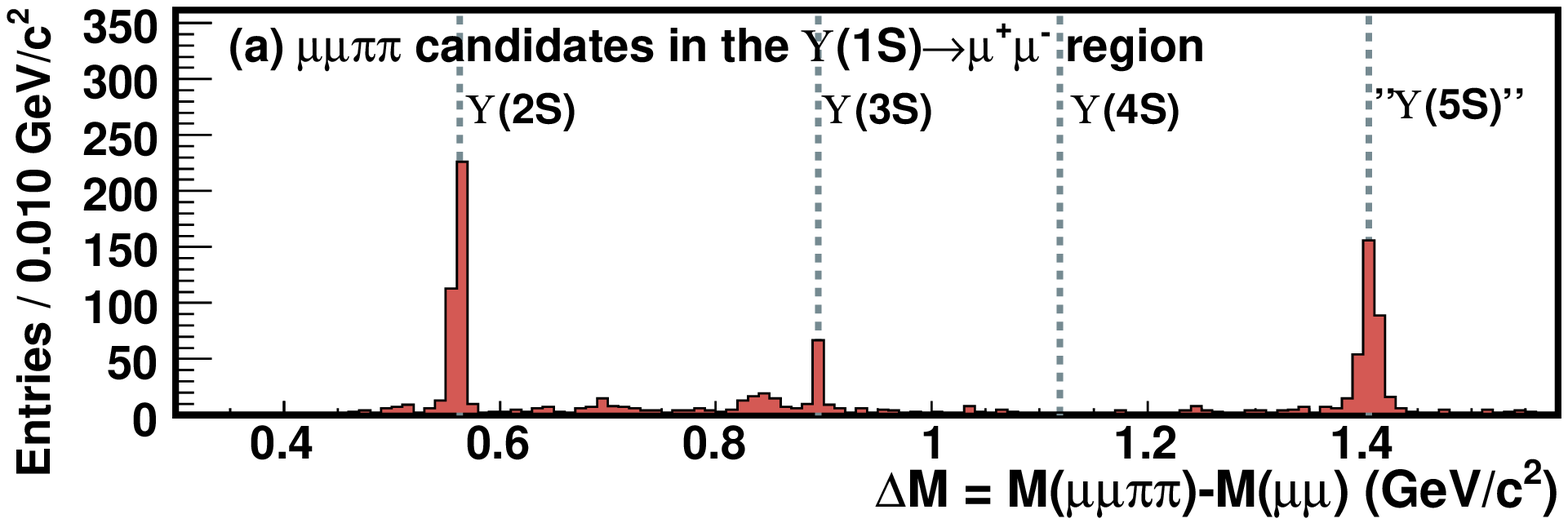}
\includegraphics[width=9.5cm]{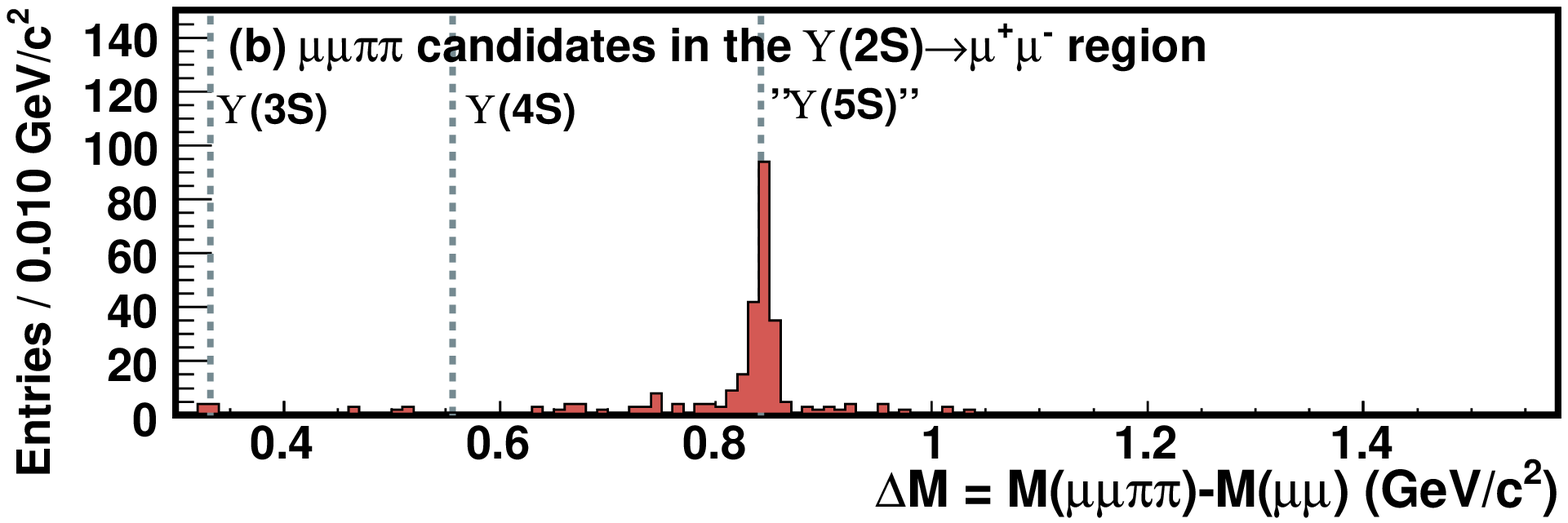}
\end{center}
\caption{The $\Delta M$ distributions for $\mu^+\mu^-\pi^+\pi^-$
events in the (a) $\Upsilon(1S)\to \mu^+\mu^-$ and (b)
$\Upsilon(2S)\to \mu^+\mu^-$ bands of Fig.~\ref{fig:ym_vs_dm}(a).
Vertical dashed lines show the expected $\Delta M$ values for 
the $\Upsilon(nS)\to\Upsilon(1,2S)\pi^+\pi^-$ transitions.}
 \label{fig:dm}
\end{figure}

The $\Delta M$
distributions for the $\mu^+\mu^-\pi^+\pi^-$ candidates in the
$\Upsilon(1S)$ and $\Upsilon(2S) \to \mu^+\mu^-$ mass bands are shown
in Fig.~\ref{fig:dm}. The peaks for
$\Upsilon(10860)\to\Upsilon(1S)\pi^+\pi^-$ and
$\Upsilon(2S)\pi^+\pi^-$ are located at $\Delta M
\sim 1.41$ GeV/$c^2$ and $\sim0.84$ GeV/$c^2$, respectively. Two other peaks at
$\Delta M \sim 0.56$ GeV/$c^2$ and $\sim0.89$ GeV/$c^2$ correspond
to $\Upsilon(2S)\to\Upsilon(1S)\pi^+\pi^-$ and
$\Upsilon(3S)\to\Upsilon(1S)\pi^+\pi^-$ transitions, respectively.
The absence of a peak around 1.12 GeV/$c^2$ corresponding to
$\Upsilon(4S)\to\Upsilon(1S)\pi^+\pi^-$ is consistent with
the rates measured in references~\cite{Aubert:2006bm,Sokolov:2006sd}. 
The structure just below $\Upsilon(3S)\to\Upsilon(1S)\pi^+\pi^-$ 
in the $\Delta M$ distribution
is from the cascade decays 
$\Upsilon(10860)\to\Upsilon(2S)\pi^+\pi^-$ with $\Upsilon(2S) \to \Upsilon(1S)[\to \mu^+\mu^-]X$.

\begin{figure}[t!]
\begin{center}
\includegraphics[width=5.5cm]{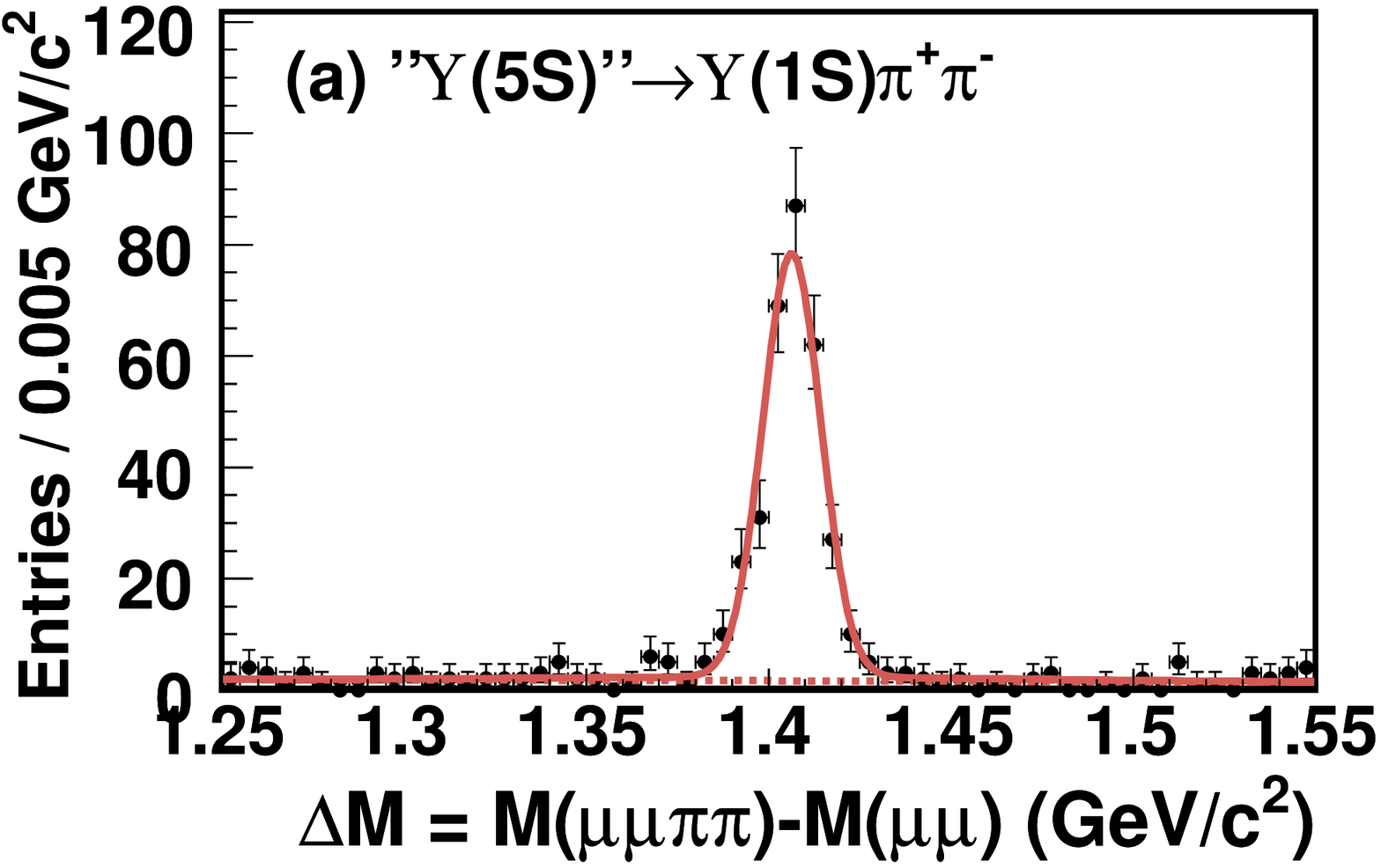}
\includegraphics[width=5.5cm]{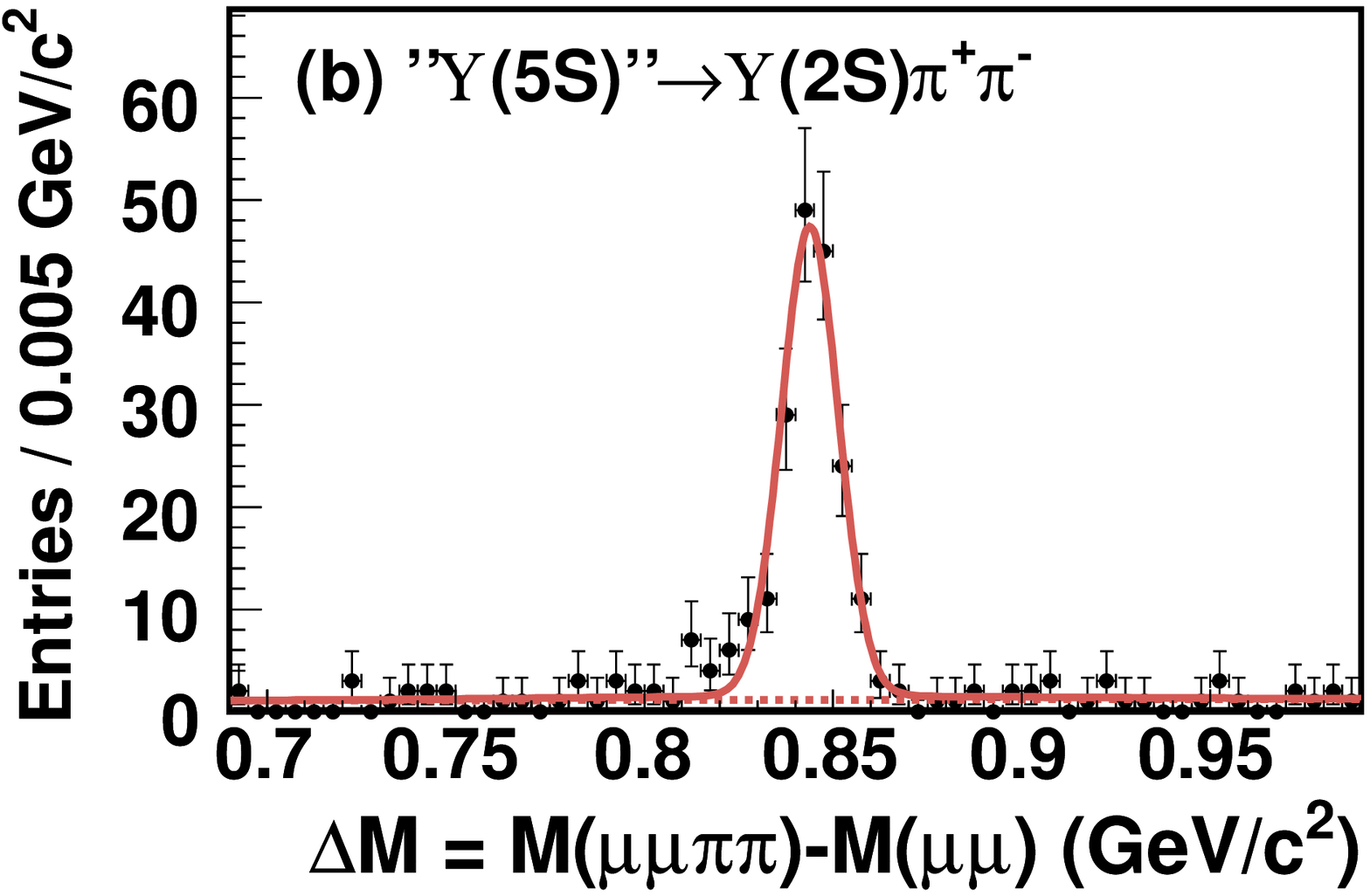}
\includegraphics[width=5.5cm]{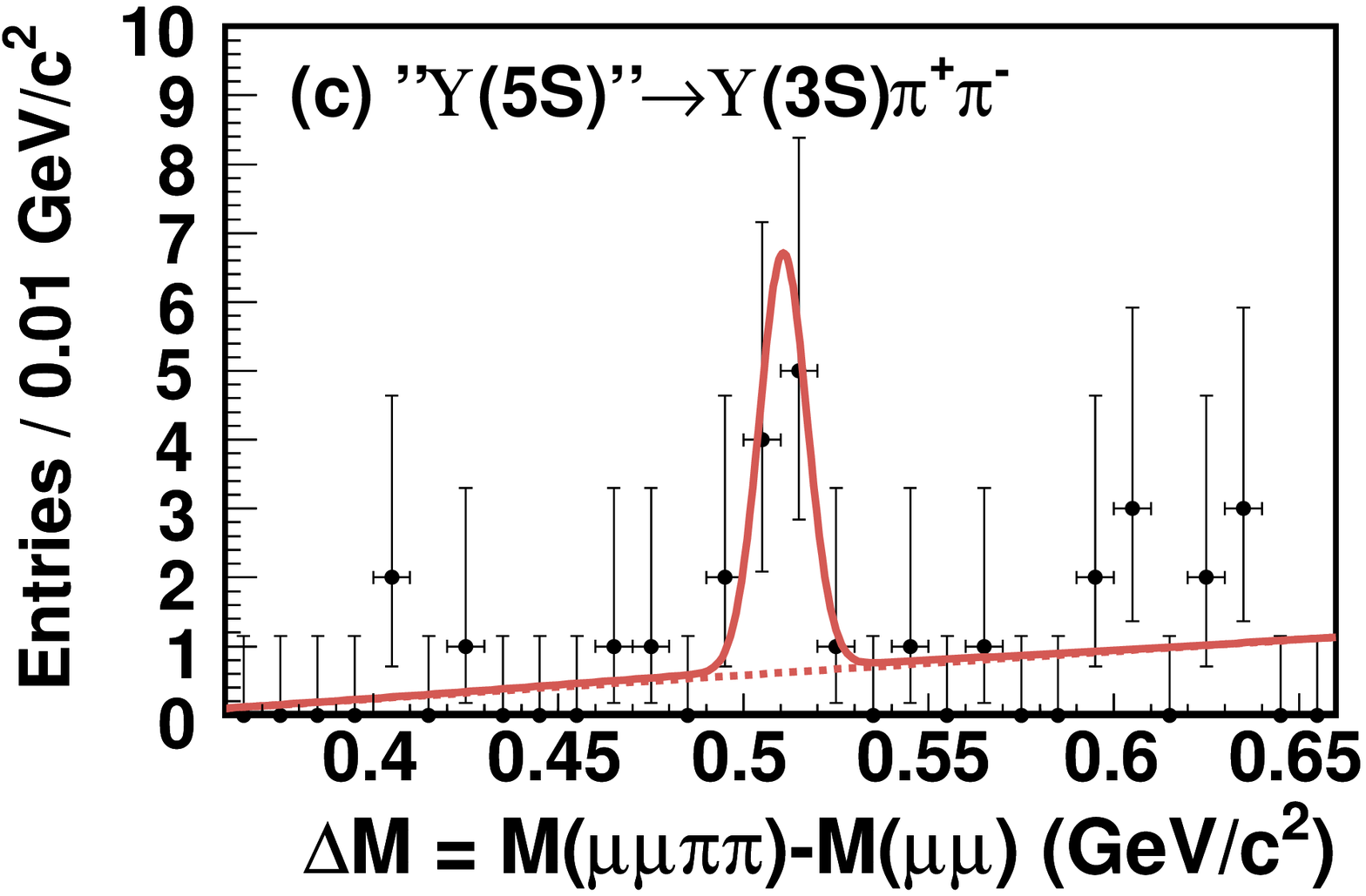}
\includegraphics[width=5.5cm]{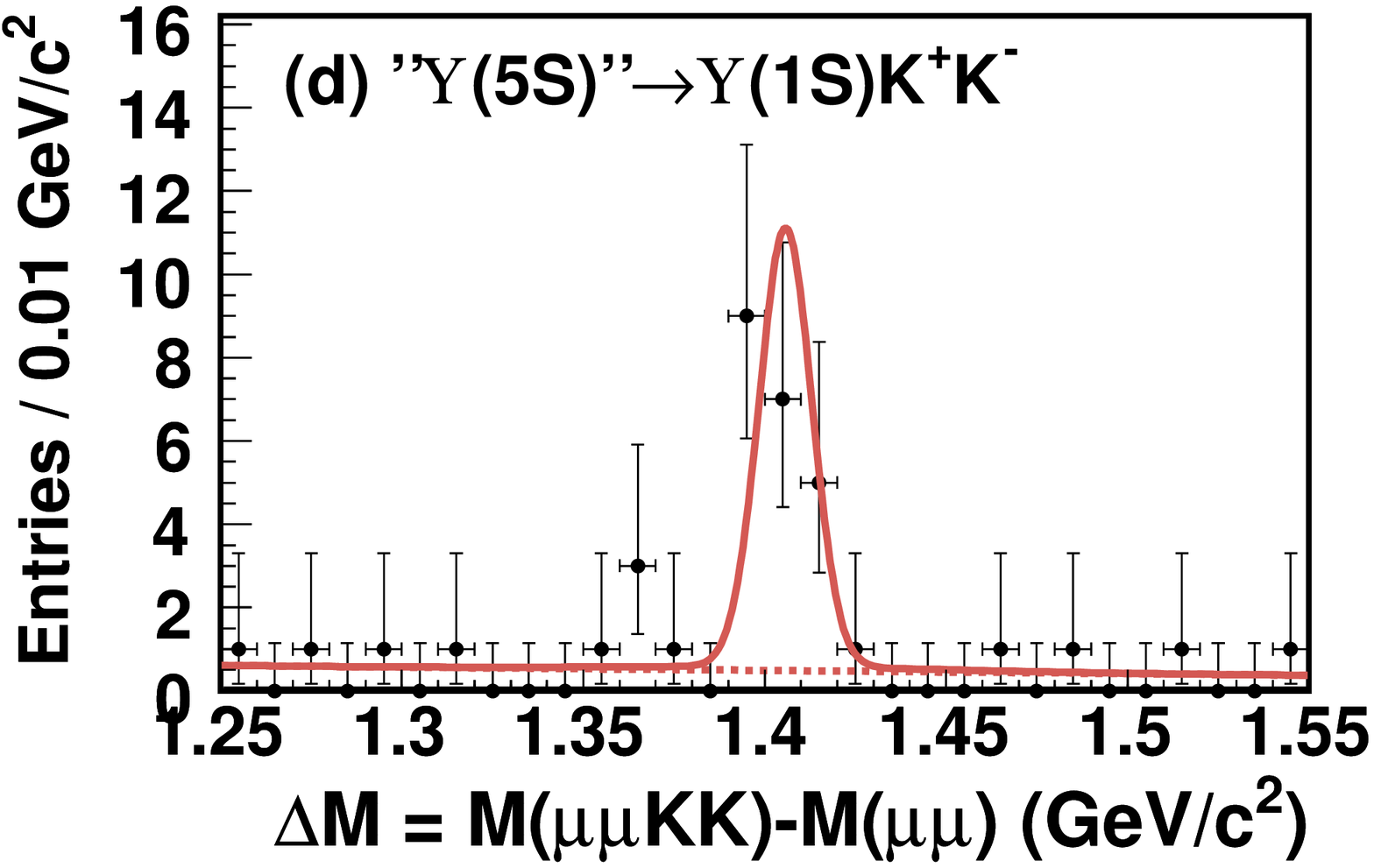}
\end{center}
\caption{The $\Delta M$ distributions for
 (a) $\Upsilon(1S)\pi^+\pi^-$,
 (b) $\Upsilon(2S)\pi^+\pi^-$,
 (c) $\Upsilon(3S)\pi^+\pi^-$, and
 (d) $\Upsilon(1S)K^+K^-$ with the fit results superimposed.
 The dashed curves show the background components in the fits.}
 \label{fig:dmfit}
\end{figure}

Signal yields are extracted by unbinned extended maximum likelihood (ML) fits to
the $\Delta M$ distributions. 
The signal is described by a sum of two Gaussians 
while the background is approximated by a linear function.
For the $\Upsilon(10860)\to \Upsilon(1S)\pi^+\pi^-$ and 
$\Upsilon(2S)\pi^+\pi^-$ modes, the remaining PDF parameters 
and yields of signal and background are floated in the fits.
For the $\Upsilon(10860)\to \Upsilon(3S)\pi^+\pi^-$ and 
$\Upsilon(1S)K^+K^-$ transitions, where statistics are limited, 
the means and widths are established based on 
$\Upsilon(10860)\to \Upsilon(1S)\pi^+\pi^-$ events and 
fixed in the fits. 
We observe $325^{+20}_{-19}$, $186\pm15$, $10.5^{+4.0}_{-3.3}$,
and $20.2^{+5.2}_{-4.5}$ events in the
$\Upsilon(10860)\to\Upsilon(1S)\pi^+\pi^-$,
$\Upsilon(2S)\pi^+\pi^-$, $\Upsilon(3S)\pi^+\pi^-$, and
$\Upsilon(1S)K^+K^-$ channels, with significances of 20$\sigma$,
14$\sigma$, 3.2$\sigma$, and 4.9$\sigma$, respectively. 
The significance
is calculated using the difference in likelihood values of the
best fit and of a null signal hypothesis including the effect of systematic uncertainties.
The Gaussian widths of the $\Upsilon(10860)\to\Upsilon(1S)\pi^+\pi^-$ and $\Upsilon(2S)\pi^+\pi^-$
peaks are found to be $8.0\pm0.5$ MeV/$c^2$ and $7.6\pm0.7$ MeV/$c^2$, respectively,
and are consistent with the MC predictions.
The distributions of $\Delta M$ with the fit results
superimposed are shown in Fig.~\ref{fig:dmfit}. 

\begin{figure}[t!]
\begin{center}
\includegraphics[width=5.5cm]{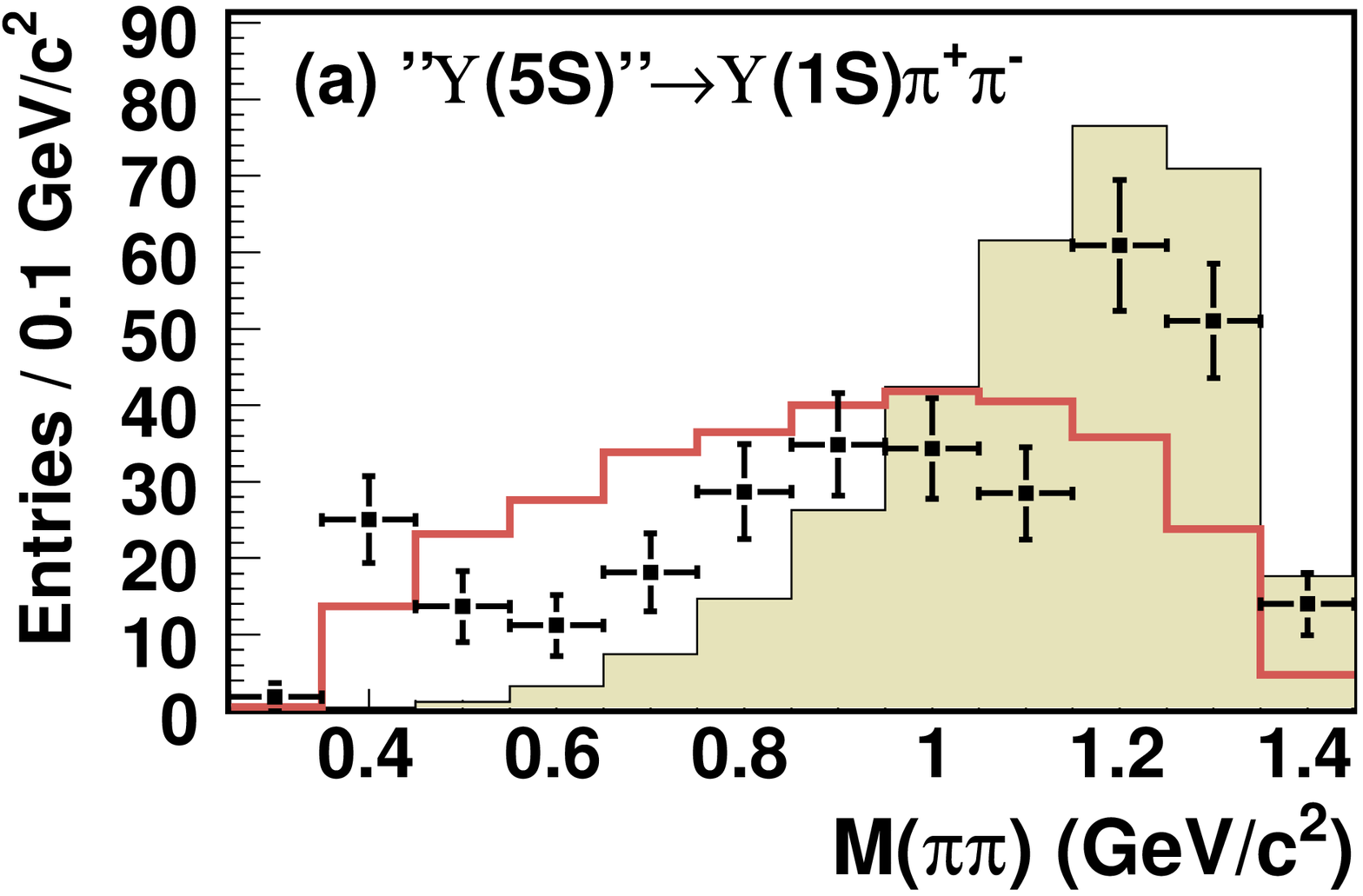}
\includegraphics[width=5.5cm]{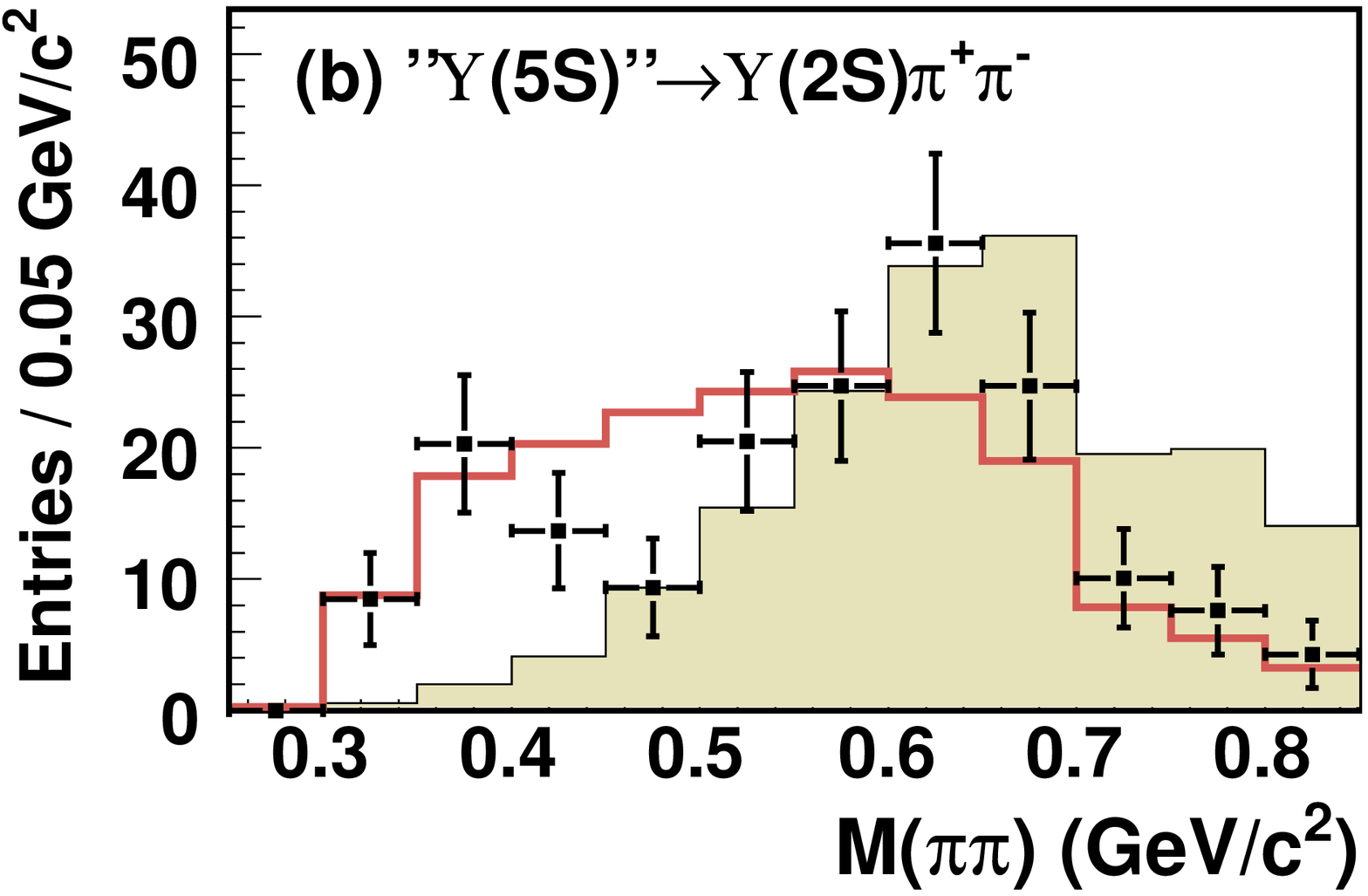}
\includegraphics[width=5.5cm]{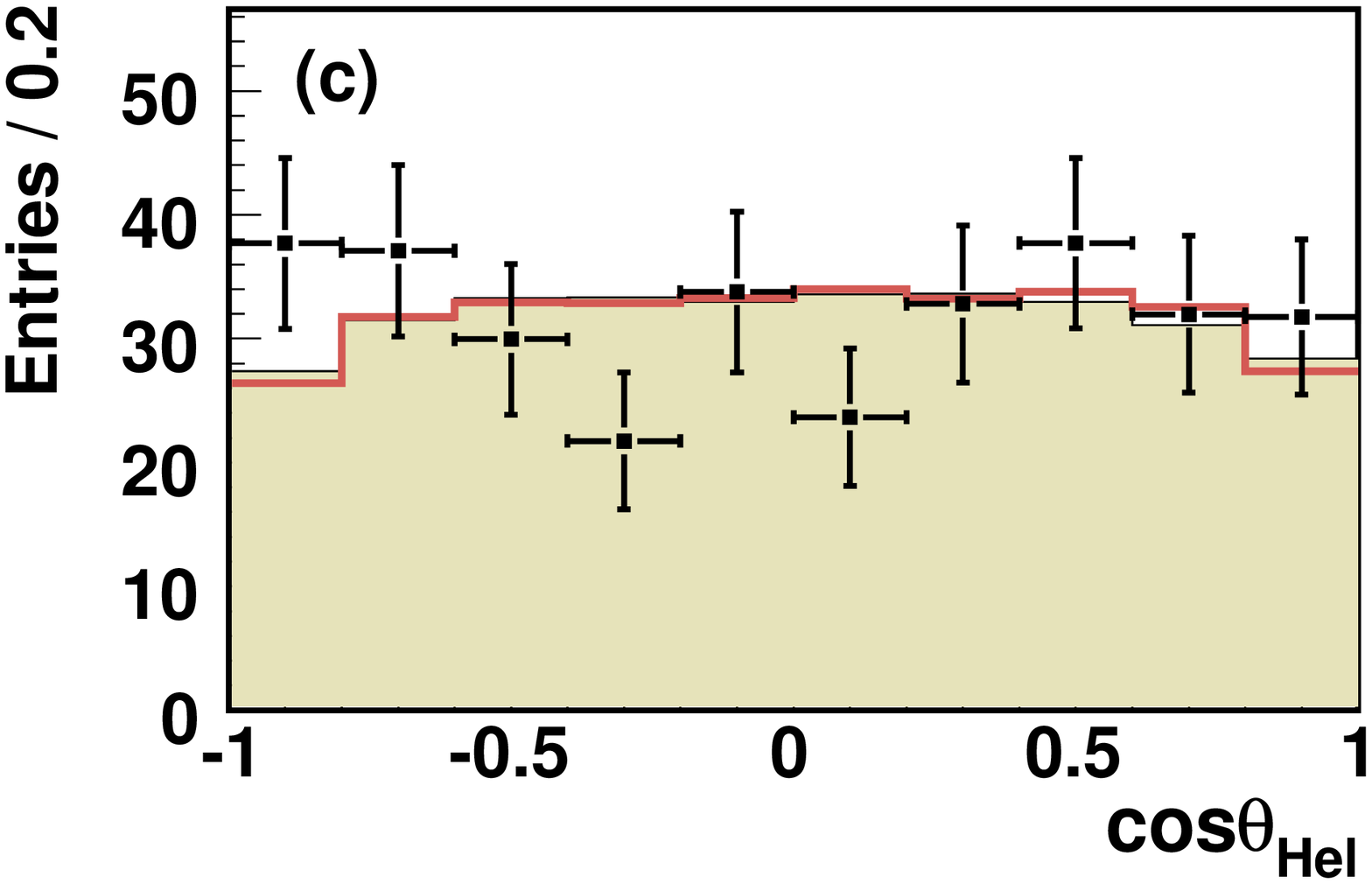}
\includegraphics[width=5.5cm]{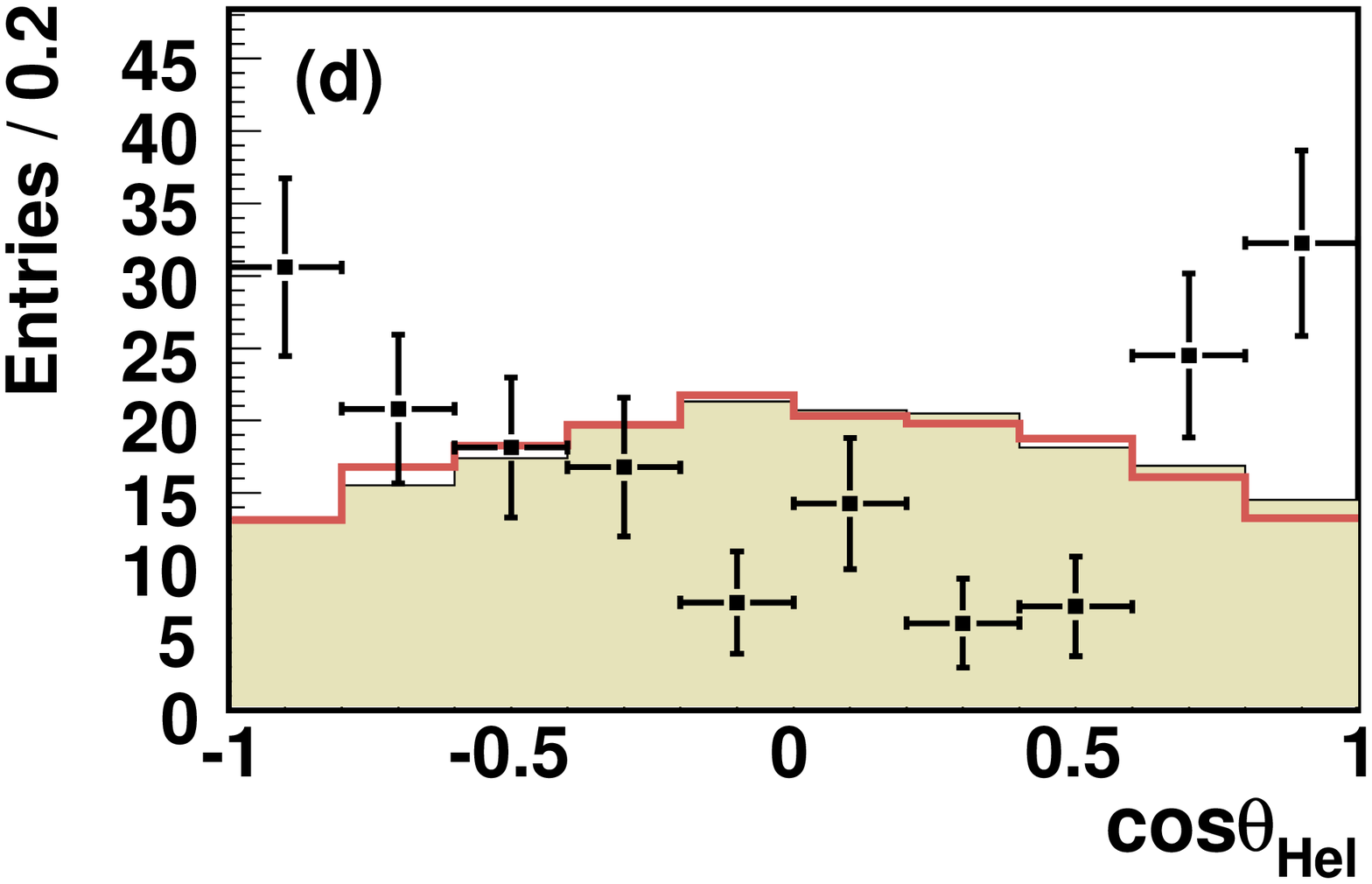}
\end{center}
\caption{The $\Upsilon(10860)$ yields as functions of $M(\pi^+\pi^-)$
and $\cos\theta_{\rm Hel}$ for (a,c) $\Upsilon(1S)\pi^+\pi^-$ and
(b,d) $\Upsilon(2S)\pi^+\pi^-$ transitions. 
The shaded (open) histogram
are from MC simulations using the model based on QCD multi-pole expansions
(phase space model).}
\label{fig:mpipi}
\end{figure}

The yields for $\Upsilon(10860)\to\Upsilon(1S)\pi^+\pi^-$,
$\Upsilon(2S)\pi^+\pi^-$ are found to be large; thus, the
corresponding invariant masses of the $\pi^+\pi^-$ system,
$M(\pi^+\pi^-)$, and the cosine of the helicity angle,
$\cos\theta_{\rm Hel}$, can be examined in detail. The helicity angle,
$\theta_{\rm Hel}$, is the angle between the $\pi^-$ and $\Upsilon(10860)$
momenta in the $\pi^+\pi^-$ rest frame.
Figure~\ref{fig:mpipi} shows the $\Upsilon(10860)$ yields as 
functions of $M(\pi^+\pi^-)$ and $\cos\theta_{\rm Hel}$, which 
are extracted using ML fits to $\Delta M$ in bins of  
$M(\pi^+\pi^-)$ or $\cos\theta_{\rm Hel}$. 
The shaded histograms in the figure are
the distributions from MC simulations using the
model in the paper~\cite{Brown:1975dz}, while the open histograms
show a  generic phase space model. 
As neither model agrees well with the observed distributions 
and the efficiencies are sensitive to both variables, 
the reconstruction
efficiencies for $\Upsilon(10860)\to\Upsilon(1S)\pi^+\pi^-$ and
$\Upsilon(2S)\pi^+\pi^-$ are obtained using MC
samples reweighted according to the measured $M(\pi^+\pi^-)$ and
$\cos\theta_{\rm Hel}$ spectra. 
Comparison of the $M(\pi^+ \pi^-)$ distribution obtained here 
with other $\Upsilon(nS)\to \Upsilon(mS) \pi^+ \pi^-$ ($m<n$) decays 
could be important for the theoretical interpretation of the results~\cite{Brown:1975dz,Simonov:2007bm}.

Assuming that signal events come only from the $\Upsilon(5S)$
resonance, the corresponding branching fractions and partial
widths can be extracted using ratios to the $\Upsilon(5S)$
cross section at $\sqrt{s}\sim10.87$ GeV, 
$0.302\pm0.015$ nb~\cite{Drutskoy:2006fg}.
The results, including the observed cross sections, are given in
Table~\ref{tab:results}. The values include the world average
branching fractions for $\Upsilon(nS)\to\mu^+\mu^-$ decays, and the total width of
the $\Upsilon(5S)$~\cite{ref:PDG2006}.  The measured partial widths, of order
$0.6$--$0.8$ MeV, are large compared to all other known 
transitions among $\Upsilon(nS)$ states. 

\begin{table*}[htbp]
\caption{Signal yield ($N_s$), significance ($\Sigma$),
reconstruction efficiency, and observed cross section ($\sigma$)
for $e^+e^- \to \Upsilon(nS)\pi^+\pi^-$ and $\Upsilon(1S)K^+K^-$
at $\sqrt{s} \sim 10.87$ GeV. Assuming the $\Upsilon(5S)$ to be the
sole source of the observed events, the branching fractions
($\mathcal{B}$) and the partial widths ($\Gamma$) for
$\Upsilon(5S) \to \Upsilon(nS)\pi^+\pi^-$ and $\Upsilon(1S)K^+K^-$
are also given. The first uncertainty is statistical, and the
second is systematic.} \label{tab:results}
\begin{center}
\footnotesize
\begin{tabular}{|lcccccc|}
\hline
Process & $N_s$ & $\Sigma$ & Eff.(\%) & $\sigma$(pb) &  $\mathcal{B}$(\%) & $\Gamma$(MeV) \\
\hline
$\Upsilon(1S)\pi^+\pi^-$ & $325^{+20}_{-19}$    &  20$\sigma$ & $37.4$  & $1.61\pm0.10\pm0.12$              &  $0.53\pm0.03\pm0.05$              & $0.59\pm0.04\pm0.09$ \\
$\Upsilon(2S)\pi^+\pi^-$ & $186\pm15$           &  14$\sigma$ & $18.9$  & $2.35\pm0.19\pm0.32$              &  $0.78\pm0.06\pm0.11$              & $0.85\pm0.07\pm0.16$ \\
$\Upsilon(3S)\pi^+\pi^-$ & $10.5^{+4.0}_{-3.3}$ & 3.2$\sigma$ &  $1.5$  & $1.44^{+0.55}_{-0.45}\pm0.19$     &  $0.48^{+0.18}_{-0.15}\pm0.07$     & $0.52^{+0.20}_{-0.17}\pm0.10$ \\
$\Upsilon(1S)K^+K^-$     & $20.2^{+5.2}_{-4.5}$ & 4.9$\sigma$ & $20.3$  & $0.185^{+0.048}_{-0.041}\pm0.028$ &  $0.061^{+0.016}_{-0.014}\pm0.010$ & $0.067^{+0.017}_{-0.015}\pm0.013$ \\
\hline
\end{tabular}
\end{center}
\end{table*}

\section{Summary}

\begin{figure}[t!]
\begin{center}
\includegraphics[width=5.5cm,height=2.2cm,angle=90]{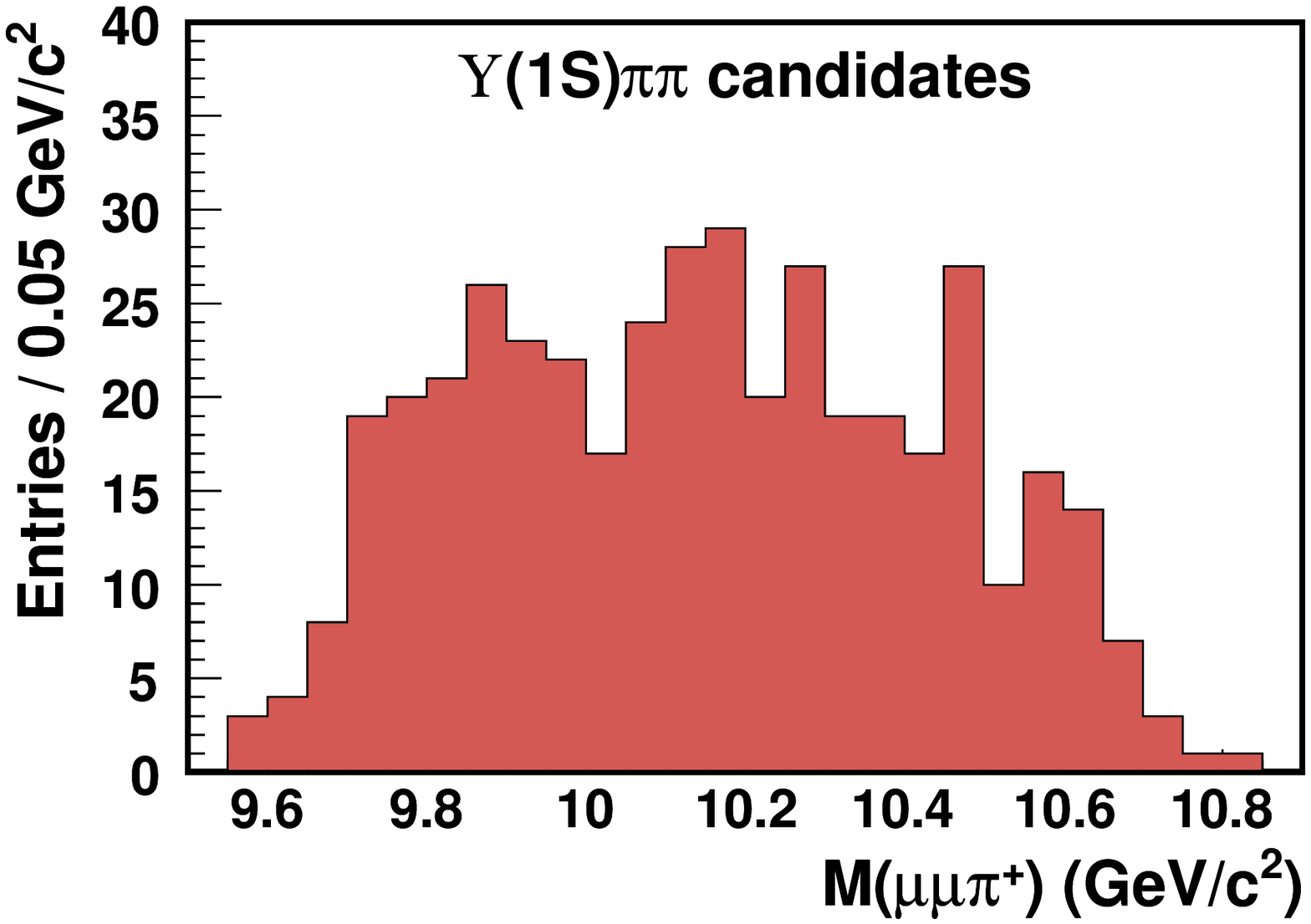}
\includegraphics[width=5.5cm,height=5.5cm]{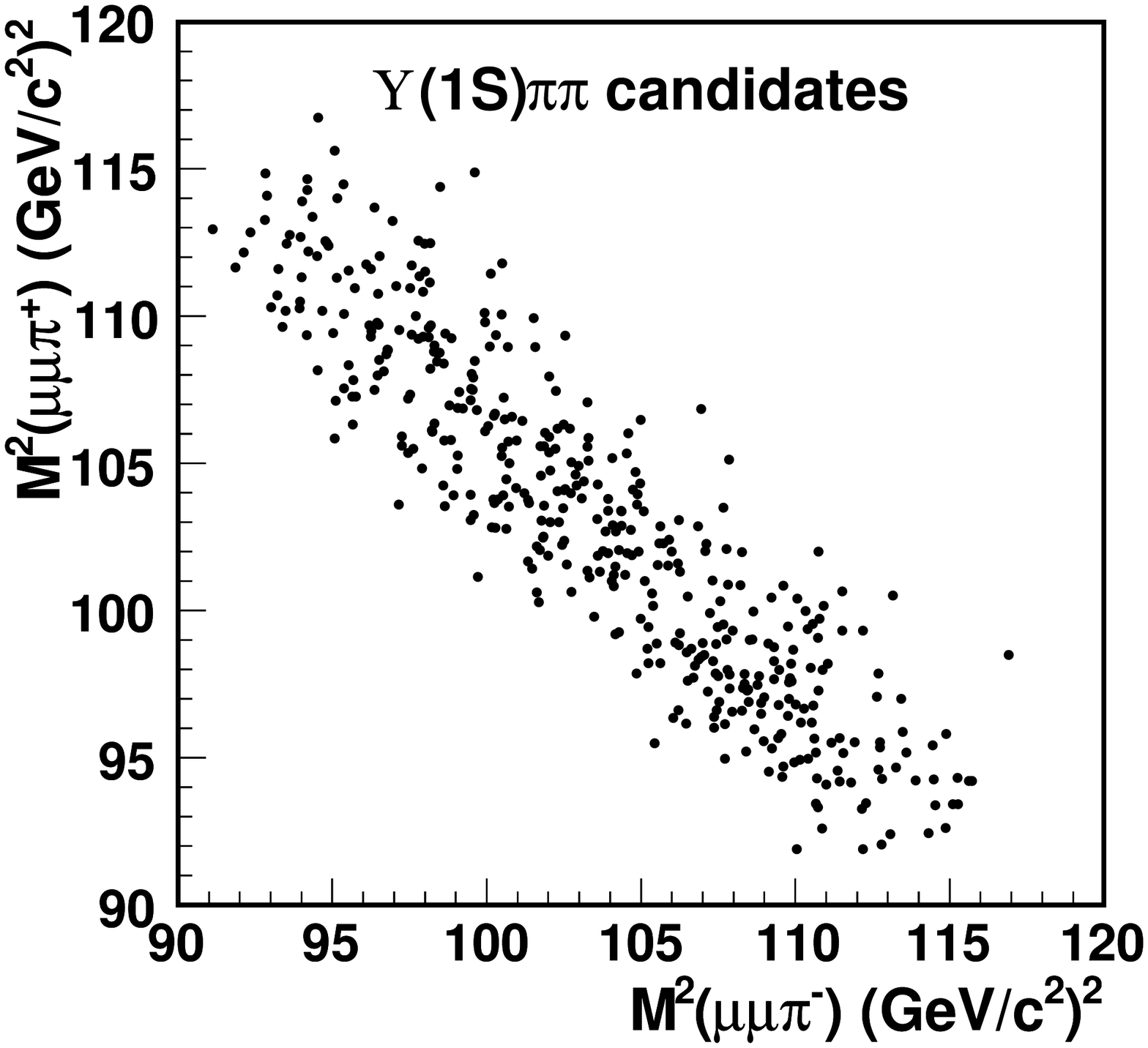}
\includegraphics[width=5.5cm,height=2.2cm,angle=90]{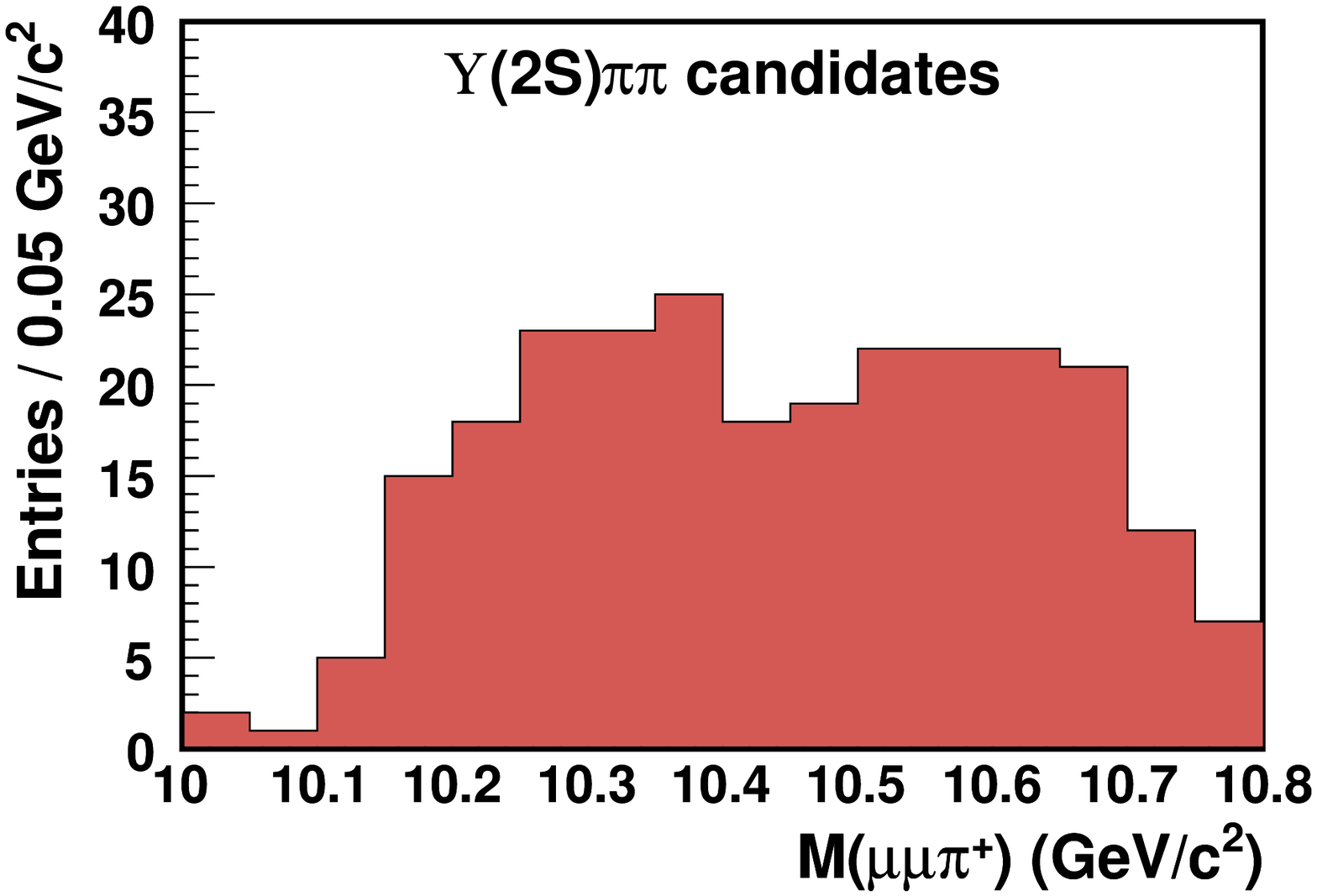}
\includegraphics[width=5.5cm,height=5.5cm]{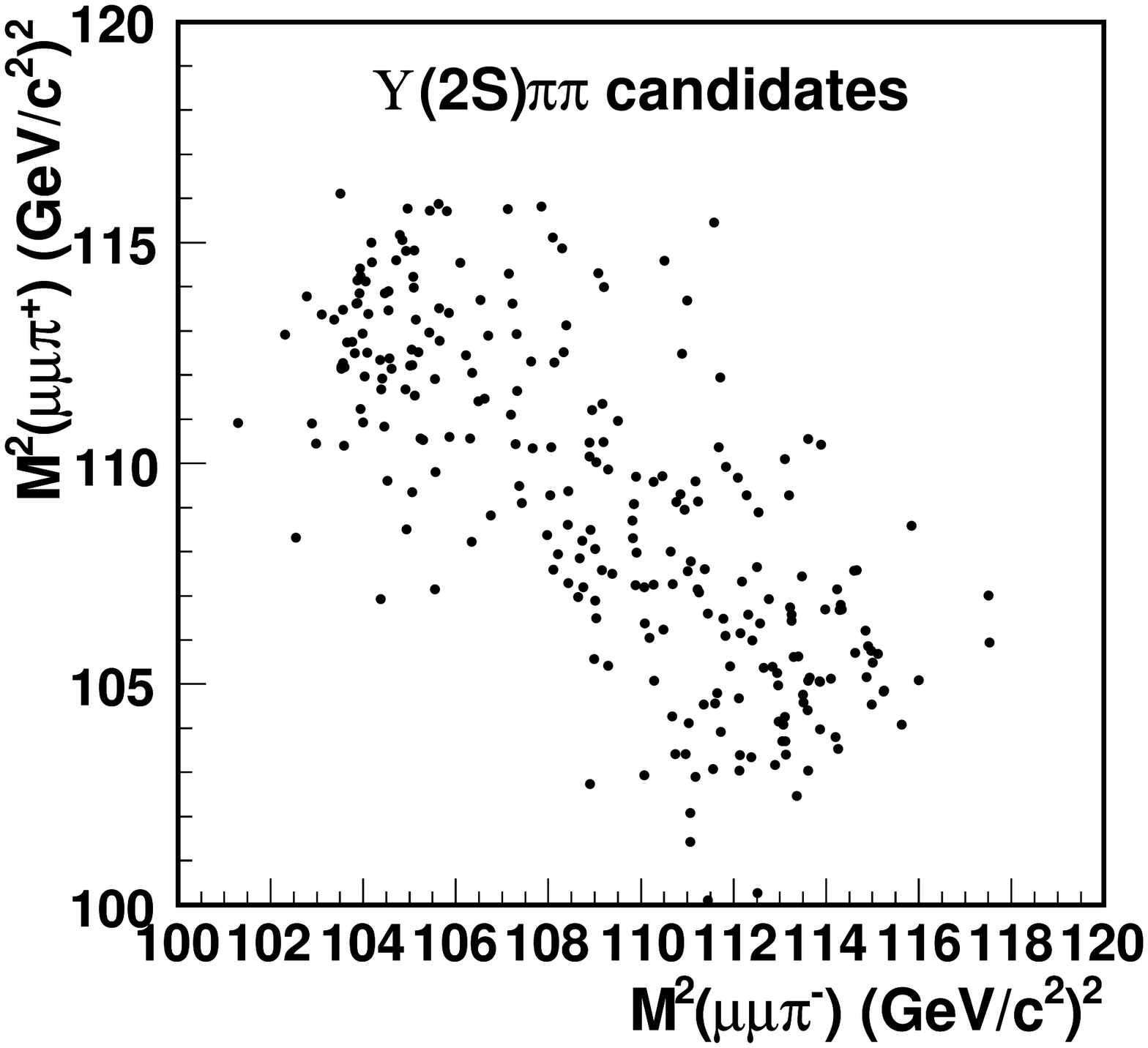}\\
\hspace{2.2cm}
\includegraphics[width=5.5cm,height=2.2cm]{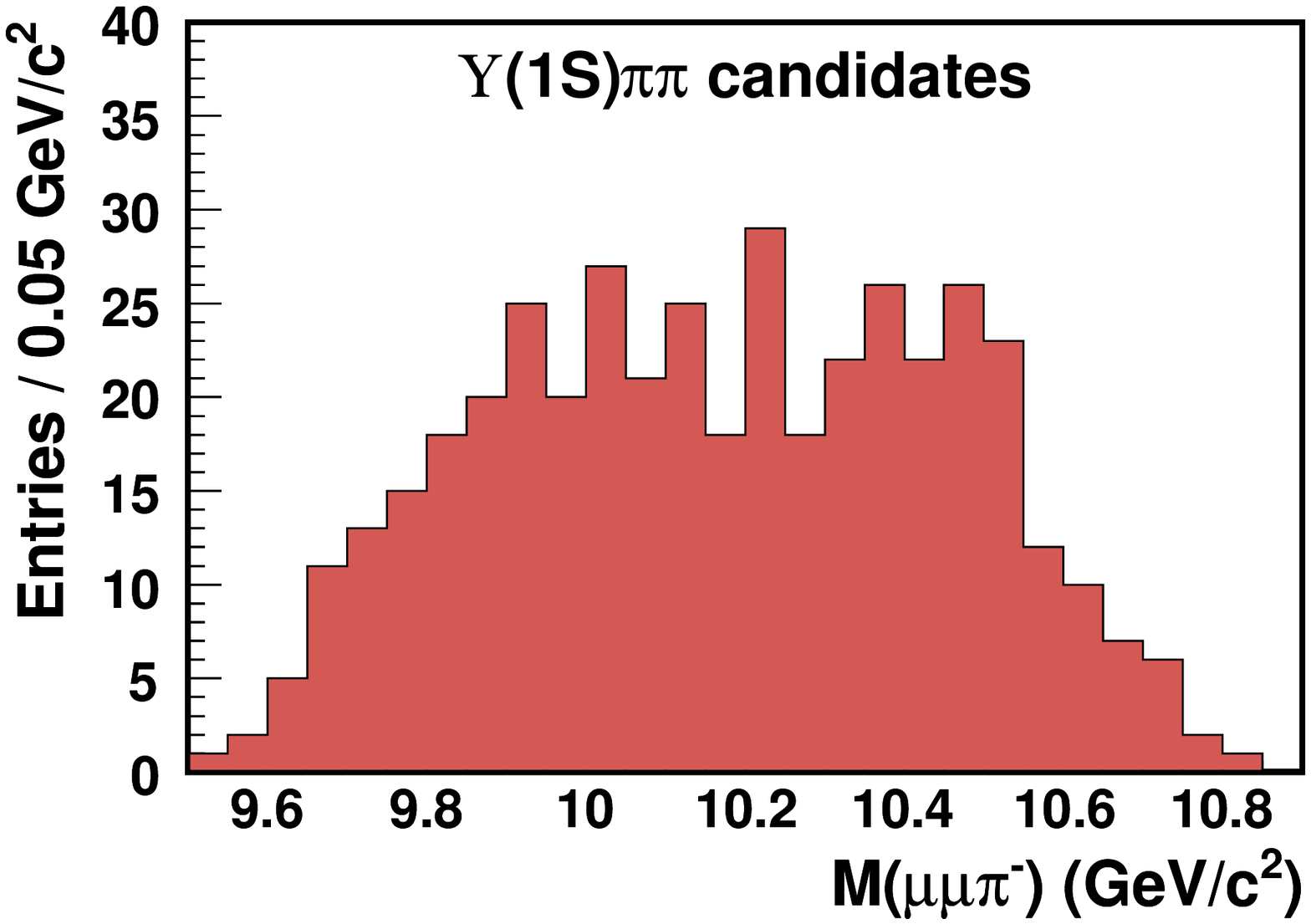}
\hspace{2.2cm}
\includegraphics[width=5.5cm,height=2.2cm]{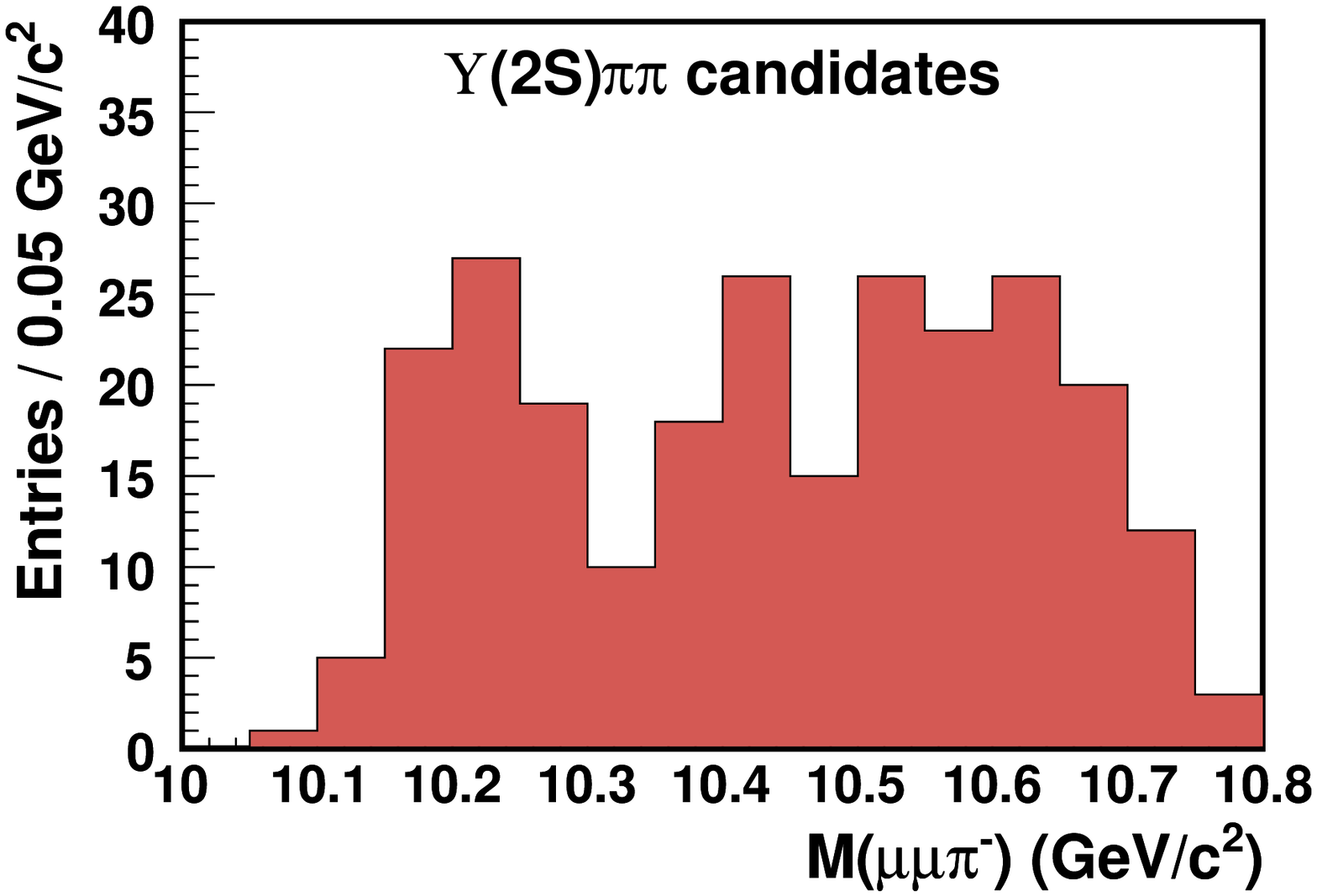}
\end{center}
\caption{The Dalitz plots, and the corresponding mass projections for $\Upsilon(1S)\pi^+\pi^-$ and
$\Upsilon(2S)\pi^+\pi^-$ candidates in the signal region. }
\label{fig:dalitz_plots}
\end{figure}

In conclusion, we report the observation of $e^+e^- \to
\Upsilon(1S)\pi^+\pi^-$ and $\Upsilon(2S)\pi^+\pi^-$ transitions, and the
evidence of $e^+e^- \to \Upsilon(3S)\pi^+\pi^-$ and
$\Upsilon(1S)K^+K^-$ transitions at a CM energy near the $\Upsilon(5S)$ resonance of
$\sqrt{s}\sim10.87$ GeV. Clear signals are observed at the expected
CM energy, with subsequent $\Upsilon(nS) \to \mu^+\mu^-$
decay. 
The measured cross sections are 
$1.61\pm0.10\pm0.12$ pb,		 
$2.35\pm0.19\pm0.32$ pb,	 
$1.44^{+0.55}_{-0.45}\pm0.19$ pb, and
$0.185^{+0.048}_{-0.041}\pm0.028$ pb for 
$e^+e^- \to \Upsilon(1S)\pi^+\pi^-$,
$\Upsilon(2S)\pi^+\pi^-$, $\Upsilon(3S)\pi^+\pi^-$, and
$\Upsilon(1S)K^+K^-$ transitions, respectively.
The first uncertainty is statistical, and the
second is systematic.
Assuming the observed signal events are due solely to the
$\Upsilon(5S)$ resonance, branching fractions are measured to be
in the range (0.48--0.78)\% for $\Upsilon(nS)\pi^+\pi^-$ channels, and 0.061\% for the $\Upsilon(1S)K^+K^-$ channel.
The corresponding
partial widths are found to be in the range (0.52--0.85) MeV for
$\Upsilon(nS)\pi^+\pi^-$, and 0.067 MeV for the $\Upsilon(1S)K^+K^-$
mode, more than two orders of magnitude
larger than the corresponding partial widths for
$\Upsilon(4S)$, $\Upsilon(3S)$ or $\Upsilon(2S)$ decays.
The unexpectedly large partial widths disagree with the
expectation for a pure $b\overline{b}$ state, unless there is a new
mechanism to enhance the decay rate, such as the existance of an intermediate resonant state~\cite{Karliner:2008rc}.
Such possibility can be examined with the Dalitz plots.
As shown in Fig.~\ref{fig:dalitz_plots}, there is no clear structure observed for the signal candidates.
A detailed energy scan within
the $\Upsilon(5S)$ energy region had been carried out at the end of 2007.
It would help to extract the resonant
spectrum, and a comparison between the yield of
$\Upsilon(nS)\pi^+\pi^-$ events and the total hadronic
cross section may help us to understand the nature of such 
signal.

\section*{Acknowledgments}

We thank the KEKB group for excellent operation of the
accelerator, the KEK cryogenics group for efficient solenoid
operations, and the KEK computer group and
the NII for valuable computing and Super-SINET network
support.  We acknowledge support from MEXT and JSPS (Japan);
ARC and DEST (Australia); NSFC (China);
DST (India); MOEHRD, KOSEF, KRF and SBS Foundation (Korea);
KBN (Poland); MES and RFAAE (Russia); ARRS (Slovenia); SNSF (Switzerland);
NSC and MOE (Taiwan); and DOE (USA).


\section*{References}
\begin{thebibliography}{99}

\bibitem{Brown:1975dz}
  L.~S.~Brown and R.~N.~Cahn,
  Phys.\ Rev.\ Lett.\  {\bf 35}, 1 (1975);
  M.~B.~Voloshin,
  JETP Lett.\  {\bf 21}, 347 (1975);
  Y.~P.~Kuang and T.~M.~Yan,
  Phys.\ Rev.\  D {\bf 24}, 2874 (1981).

\bibitem{CLEO:2007sja}
  F.~Butler {\it et al.} [CLEO Collaboration], 
  Phys.\ Rev.\ D {\bf 49}, 40 (1994);
  D.~Cronin-Hennessy {\it et al.}  [CLEO Collaboration],
  Phys.\ Rev.\  D {\bf 76}, 072001 (2007).

\bibitem{Aubert:2006bm}
  B.~Aubert {\it et al.}  [BaBar Collaboration],
  Phys.\ Rev.\ Lett.\  {\bf 96}, 232001 (2006).

\bibitem{Sokolov:2006sd}
  A.~Sokolov {\it et al.}  [Belle Collaboration],
  Phys.\ Rev.\  D {\bf 75}, 071103 (2007).
  
\bibitem{Simonov:2007bm}
  Yu.~A.~Simonov,
  JETP\ Lett.\ {\bf 87}, 147 (2008).

%
\bibitem{Y4260}
  B.~Aubert {\it et al.}  [BaBar Collaboration],
  Phys.\ Rev.\ Lett.\  {\bf 95}, 142001 (2005).

\bibitem{Hou:2006it}
  W.~S.~Hou,
  Phys.\ Rev.\  D {\bf 74}, 017504 (2006).

\bibitem{Abe:2007tk}
  K.~F.~Chen {\it et al.}  [Belle Collaboration],
  Phys.\ Rev.\  Lett.\  {\bf 100}, 112001 (2008).


\bibitem{ref:PDG2006} {W.-M.~Yao {\it et al.}, J. Phys. G {\bf 33}, 1 (2006).}

\bibitem{Drutskoy:2006fg}
  A.~Drutskoy {\it et al.}  [Belle Collaboration],
  Phys.\ Rev.\ Lett.\  {\bf 98}, 052001 (2007).

\bibitem{Karliner:2008rc}
  M.~Karliner and H.~J.~Lipkin,
  arXiv:0802.0649 [hep-ph].


\end{thebibliography}
\end{document}